\documentclass[journal]{IEEEtran}
\usepackage[T1]{fontenc}
\usepackage[utf8]{inputenc}
\usepackage{amsmath,amssymb,bm,booktabs,array,tabularx,multirow,cite}
\usepackage{graphicx,xcolor,tikz,pgfplots,pgfplotstable}
\usepackage{algpseudocode}
\usepackage[hidelinks,breaklinks]{hyperref}
\usetikzlibrary{arrows.meta,positioning,fit}
\usepgfplotslibrary{fillbetween,groupplots}
\pgfplotsset{compat=1.18}
\pdfmapfile{=vendor/ieee-fonts.map}
\newcommand{\R}{\mathbb{R}}

\newcommand{\diag}{\operatorname{diag}}

\newcommand{\argmin}{\operatorname*{arg\,min}}
\newcounter{algorithm}
\newenvironment{algorithm}[1]{\begin{figure}[t]\refstepcounter{algorithm}\hrule\smallskip\noindent\textbf{Algorithm \thealgorithm: #1}\par\smallskip\small}{\smallskip\hrule\end{figure}}
\newcommand{\TwinRMSE}{0.1261}
\newcommand{\TwinCoverage}{90.3}
\newcommand{\TwinBrier}{0.2003}

\newcommand{\EWMARMSE}{0.1452}

\newcommand{\EWMABrier}{0.2005}

\newcommand{\UncorrectedCoverage}{63.1}

\newcommand{\NoSurveyRMSE}{0.1410}

\newcommand{\NoSurveyBrier}{0.2004}

\newcommand{\DiagonalRCoverage}{86.2}

\newcommand{\RelativeRMSEGain}{13.1}
\newcommand{\PairedDifference}{-0.0191}
\newcommand{\PairedDifferenceLow}{-0.0213}
\newcommand{\PairedDifferenceHigh}{-0.0169}
\newcommand{\PairedHolmP}{0.0137}
\newcommand{\ShiftRMSE}{0.1559}
\newcommand{\ShiftStateCoverage}{82.9}
\newcommand{\ShiftGaussianCoverage}{75.4}
\newcommand{\ShiftGaussianScore}{2.452}

\newcommand{\ShiftACICoverage}{84.1}
\newcommand{\ShiftACIScore}{2.185}
\newcommand{\ShiftPIDCoverage}{83.5}
\newcommand{\ShiftPIDScore}{2.188}
\newcommand{\OutageRMSE}{0.1407}
\newcommand{\OutageStateCoverage}{90.5}

\newcommand{\MnarRMSE}{0.1339}
\newcommand{\MnarStateCoverage}{90.4}

\newcommand{\MeanRegret}{0.115}

\newcommand{\ConservativeRegret}{0.293}

\newcommand{\ConservativeDeferred}{6.3}

\newcommand{\RandomRegret}{0.248}

\newcommand{\CILogisticBrier}{0.1289}

\newcommand{\CIRandomForestBrier}{0.1293}

\newcommand{\Seeds}{10}
\newcommand{\Worlds}{40}

\newcommand{\Weeks}{104}
\newcommand{\OrgCount}{48}
\newcommand{\TrainOrgs}{24}
\newcommand{\ValOrgs}{12}
\newcommand{\TestOrgs}{12}

\newcommand{\CITestRows}{13\,074}
\newcommand{\CIUsedTrainRows}{75\,000}

\newcommand{\CITestProjects}{224}
\newcommand{\CIRawJobs}{2\,640\,825}
\newcommand{\CIEligibleBuilds}{357\,607}
\newcommand{\ElapsedSeconds}{34.9}
\newcommand{\PeakRSS}{267.0}

\newcommand{\SourceCorrelation}{0.65}

\title{ Bayesian Evidence Fusion for Capability Tracking---A Simulation Study and Historical CI Pilot}
\author{Taimoor Ahmad\thanks{Superior University, Lahore Pakistan.}}
\begin{document}
\maketitle
\begin{abstract}
SaaS capability assessments can become stale between questionnaires, while delivery telemetry offers incomplete and correlated evidence. Continuous assessment and Bayesian delivery analysis already exist, leaving uncertainty about how biased surveys and missing observations should influence capability estimates. We implement SAAS-MATURITY-TWIN, an anchored Gaussian state model that jointly filters capability proxies and persistent survey offsets. Source covariance prevents correlated observations from being counted independently. Conditional uncertainty supports forecasting and budgeted intervention screening, with action effects estimated from randomized synthetic cohorts. Ten seeds generate forty paired worlds spanning reference observations, extensive missingness, informative missingness, and increased noise. Reference latent root mean squared error is \TwinRMSE{}, compared with \EWMARMSE{} for a tuned moving average; nominal ninety percent intervals cover \TwinCoverage{} percent. Improved state recovery produces only small incident forecasting differences. Adaptive conformal inference and an adapted conformal PID controller are evaluated separately for observed telemetry, where noise shifts expose calibration limitations. A licensed TravisTorrent pilot evaluates build prediction on \CITestRows{} held out builds from \CITestProjects{} projects. Logistic regression achieves a Brier score of \CILogisticBrier{}, close to random forests. These results support reproducible methodological testing, but do not establish real organizational maturity, intervention effectiveness, or production readiness. Independent calibration and longitudinal validation remain necessary.

\end{abstract}
\section{Introduction}\label{sec:introduction}
Software as a service (SaaS) organizations make recurring decisions about deployment safety, incident recovery, security automation, and engineering investment. A capability assessment can help structure these decisions, but its usefulness depends on what the measurements mean and how quickly they become outdated. A team may report that rollback is automated while recent deployments show repeated manual recovery. Another team may introduce tracing without seeing an immediate improvement in incident outcomes. Both situations require an assessment that distinguishes evidence about a capability from evidence about its operational consequences. A maturity label alone does not supply that distinction.

The SaaS Application Maturity Assessment Model organizes questionnaire-based assessment around design, architecture, business performance, and the overall organization \cite{Aleem2024SaaS,alkobaisi2025artificial}. It supplies a domain-specific starting point rather than a reason to infer those four dimensions from repository activity. DevOps capability instruments likewise provide explicit practices and measurement procedures \cite{Plant2022Instrument}. The broader maturity literature includes many assessment approaches \cite{Jayakody2024Review}, while research has cautioned that pursuing maturity processes is not sufficient to secure organizational performance \cite{Marnewick2021Maturity}. A technically active project can still have poor customer outcomes, and an organization with sparse public telemetry can have strong internal controls. Missing evidence therefore cannot be interpreted automatically as low maturity.

Continuous assessment itself is established prior work. The Continuous Maturity Assessment Method describes iterative tracking of organizational capabilities and has been evaluated in organizational settings \cite{Stoiber2023Continuous}. Capability--metric mappings \cite{Amaro2023Capabilities}, a systematic capability taxonomy \cite{Amaro2023Practices}, and metric syntheses \cite{Amaro2024Metrics} already connect improvement practices with measurable evidence. Automated DORA-oriented studies extract longitudinal project metrics \cite{Wilkes2023DORA,yu2007tornado} and integrate production-adjacent tools for continuous measurement \cite{Ruegger2024Automated}. More recently, BayesInsights applies Bayesian networks to software delivery and developer experience in an industrial organization \cite{Kirbas2026BayesInsights,alkobaisi2012interactive}. Consequently, neither adding telemetry to a maturity questionnaire nor applying a Bayesian model is, by itself, a defensible new contribution.

The narrower problem examined here is measurement under disagreement. Two dashboards can count overlapping events, a survey can contain a persistent offset, and a telemetry source can disappear during a difficult period. An equally weighted score treats these situations as changes in capability even when they are changes in measurement. A conventional Gaussian filter supplies temporal smoothing, but a filter that treats a biased survey as an unbiased observation can become confidently wrong. Similarly, assuming independent sources can narrow an uncertainty interval without adding equivalent information. These issues motivate a source-aware measurement model, not a claim that an unobserved organizational construct has already been identified.

We represent three technical capability proxies by a continuous latent state and treat an aggregate questionnaire score as one noisy observation channel. The proposed implementation augments that state with an organization-specific survey offset and updates it using the covariance of the available telemetry sources. It retains a prior when all evidence is absent and reports uncertainty conditional on an externally calibrated measurement model. The coordinate system is anchored by fixed telemetry loadings and an agreed polarity. That anchoring is an assumption: it does not establish that a unit of the synthetic state corresponds to an empirically meaningful change in engineering capability. This distinction follows the broader measurement-validity problem discussed by Borsboom et al. \cite{borsboom2004}.

Data access sets the evidential scope. No partner dataset joining validated surveys, deployment events, incidents, costs, and intervention histories was supplied. The audited public sources offer components of that evidence but not a verified joined capability ground truth. We therefore conduct a labeled simulation study, where latent recovery and policy regret are observable by construction. A separate licensed TravisTorrent pilot tests whether a simple project-disjoint, forward-time CI workflow can be executed on real records \cite{beller2017,ahmed2017open}. It neither transfers the simulator's latent labels to real projects nor treats ordinary OSS projects as SaaS organizations. This separation makes the public pilot useful as a data-engineering check while preventing it from being mistaken for construct validation.

Three questions guide the study. RQ1 asks whether explicitly modeling a persistent survey offset and correlated sources improves latent recovery and conditional uncertainty in the chosen simulator. RQ2 asks how missingness and observation noise shifts change state and telemetry-forecast calibration. RQ3 asks how estimated action effects and uncertainty influence a constrained recommendation relative to an oracle defined by that simulator. The first two questions test a measurement implementation; the third tests a decision procedure under known structural assumptions. None estimates the causal benefit of a real production intervention.

The experiments use disjoint synthetic training, validation, and test organizations. Model selection uses preintervention observed-telemetry likelihood rather than test latent labels. Ten independent seeds generate paired observation regimes. In the reference regime, the full model has RMSE \TwinRMSE{}, compared with \EWMARMSE{} for a validation-tuned exponentially weighted moving average (EWMA). Its nominal 90\% state interval covers \TwinCoverage{}\% of synthetic states. These differences are substantial for the simulator's latent-recovery task, but the corresponding incident Brier scores, \TwinBrier{} and \EWMABrier{}, differ only slightly. The public CI pilot similarly offers no reason to favor model complexity automatically: logistic regression is close to random forests. These mixed observations constrain the practical interpretation.

The specific contributions are:
\begin{itemize}
\item An executable, covariance-aware survey--telemetry fusion prototype with a persistent survey-offset state, explicit evidence provenance, missing-source updates, and conditional credible intervals. The filtering equations are established methods; the contribution is their inspectable assessment integration.
\item A paired simulation protocol separating latent recovery, observed-telemetry interval calibration, and downstream prediction, including biased surveys, source dependence, informative missingness, and noise shifts. Recent ACI and PID comparators are evaluated only for their appropriate observation-forecast task.
\item An auditable budgeted action-selection experiment using randomized synthetic contrasts and a separate historical CI feasibility pilot. Both include negative or neutral findings, executed results, and a documented boundary on organizational and causal claims.
\end{itemize}

The remainder of this paper is organized as follows. Section~\ref{sec:related} positions the work against maturity assessment, telemetry, and statistical methods. Section~\ref{sec:model} defines the measurement model. Section~\ref{sec:method} presents filtering and intervention screening. Section~\ref{sec:experiments} specifies the experiments, and Section~\ref{sec:results} reports their results. Section~\ref{sec:limitations} discusses validity and submission limitations. Section~\ref{sec:conclusion} concludes.

\section{Related Work}\label{sec:related}
\subsection{Capability instruments and continuous assessment}
Questionnaire models specify what should be measured, while telemetry systems specify what can be observed repeatedly. These roles overlap but are not interchangeable. The SaaS model \cite{Aleem2024SaaS} and the participatory capability instrument \cite{Plant2022Instrument} motivate retaining structured assessment evidence. DevOps capability and practice reviews \cite{Amaro2023Practices,Amaro2025Mapping} provide a broader taxonomy, but selecting three technical proxies from it does not validate a replacement instrument. Context also matters: Plant et al. examine context--capability coalignment in public-sector organizations \cite{Plant2025Coalignment,bae2007interactive}. Pooling organizations without accounting for such differences can conceal rather than resolve measurement problems.

Stoiber et al. \cite{Stoiber2023Continuous} are particularly close to the original continuous-assessment proposal. Their work means that temporal tracking cannot serve as the novelty claim here. Amaro et al. \cite{Amaro2023Capabilities} further connect capabilities to outcome metrics, while Marnewick and Langerman \cite{Marnewick2021Maturity} question the assumption that maturity progression alone secures performance. The unresolved implementation question pursued in this paper is how an explicitly uncertain measurement process should respond when the sources disagree. Our simulator can examine the statistical consequences of that design choice; it cannot demonstrate that an actual capability has been measured validly.

\subsection{Automated telemetry and Bayesian decision support}
Wilkes et al. \cite{Wilkes2023DORA} automate operationalizations of historical DORA metrics across public repositories. R\"uegger et al. \cite{Ruegger2024Automated} demonstrate automated measurement in an industrial microservice setting. Their contribution precedes the proposed evidence ingestion pipeline. The metric review by Amaro et al. \cite{Amaro2024Metrics} also establishes that definitions and collection procedures are themselves substantial concerns. We therefore avoid rebranding build failure as change failure, or build duration as incident recovery. The CI pilot has deliberately narrower labels.

BayesInsights \cite{Kirbas2026BayesInsights} is a second important novelty constraint. It models dependencies among software delivery and developer-experience factors with Bayesian networks. Its existence rules out a general claim that Bayesian software-delivery analysis is new. Its internal source data and implementation were not publicly reproducible in the sources inspected, so we do not label a generic Bayesian network as its reproduced baseline. The present experiment instead exposes the state and observation assumptions of a small dynamic measurement model. Table~\ref{tab:closest} summarizes thirteen directly relevant studies; numerical findings from those studies are not used as experimental comparison scores.

\begin{table*}[t]
\caption{Closest domain studies and the present scope. The final column identifies what the sources inspected do not establish for this project; it is not a comprehensive absence claim about each paper. No published performance value is mixed with our results.}\label{tab:closest}
\centering\small
\begin{tabularx}{\textwidth}{p{2.35cm}p{2.9cm}Xp{4.35cm}}
\toprule
Study & Setting or evidence & Established focus & Remaining requirement for this study\\
\midrule
Aleem et al. \cite{Aleem2024SaaS}, 2024 & SaaS assessment model & Questionnaire, dimensions, rating & Joint dynamic evidence and independent outcome validity\\
Jayakody et al. \cite{Jayakody2024Review}, 2024 & Literature review & DevOps maturity review & Primary-content details not fully accessible here\\
Plant et al. \cite{Plant2022Instrument}, 2022 & Expert and team field input & Capability measurement instrument & Repeated telemetry and uncertainty validation\\
Amaro et al. \cite{Amaro2023Practices}, 2023 & Multivocal review & Capabilities and practices & Calibrated observation-to-capability mapping\\
Amaro et al. \cite{Amaro2023Capabilities}, 2023 & Design science and interviews & Capability--metric relationships & Dynamic bias-aware measurement\\
Amaro et al. \cite{Amaro2024Metrics}, 2024 & Multivocal metric review & Metrics and KPI definitions & Target-setting measurement invariance\\
Stoiber et al. \cite{Stoiber2023Continuous}, 2023 & Organizational evaluations & Continuous maturity assessment & Explicit probabilistic source disagreement\\
Wilkes et al. \cite{Wilkes2023DORA}, 2023 & Public repositories & Automated historical DORA proxies & Production semantics and latent validity\\
R\"uegger et al. \cite{Ruegger2024Automated}, 2024 & Industrial microservices & Automated continuous metrics & Joint survey/telemetry uncertainty study\\
Plant et al. \cite{Plant2025Coalignment}, 2025 & Public-sector DevOps & Context--capability coalignment & Cross-context longitudinal validation\\
Marnewick et al. \cite{Marnewick2021Maturity}, 2021 & Organizational performance & Limits of maturity pursuit & Causal outcome evidence for an intervention\\
Amaro et al. \cite{Amaro2025Mapping}, 2025 & Adoption mapping study & Capabilities across stages & Source-aware dynamic estimation\\
Kirbas et al. \cite{Kirbas2026BayesInsights}, 2026 & Industrial Bayesian analysis & Delivery/developer-experience dependencies & Public dynamic evidence-fusion comparison\\
\textbf{This work} & \textbf{Simulation; historical CI pilot} & \textbf{Bias-aware dynamic fusion and stress protocol} & \textbf{Real SaaS construct and intervention validation remain open}\\
\bottomrule
\end{tabularx}
\end{table*}

\subsection{Uncertainty, missing data, and intervention effects}
The recursive estimator uses Kalman's established linear filtering framework \cite{kalman1960}. Proper scoring rules \cite{gneiting2007} and interval scores \cite{bracher2021} motivate assessing calibration together with width, rather than rewarding broad intervals solely for coverage. The Bracher paper has a published correction affecting examples and figure values; we implement the score definition directly rather than copying those examples. ACI \cite{aci2021} and conformal PID control \cite{pid2023} supply recent sequential observation-interval comparators. The later online conformal framework \cite{gibbs2024} further highlights that adaptation rules and local guarantees differ. These methods are not latent maturity estimators, so their comparisons occupy a separate task.

Ignoring a missing observation is justified only under assumptions on the observation process \cite{rubin1976}. Informative telemetry loss can violate those assumptions, and even parametric nonignorable missing-data models have identification limitations \cite{miao2016}. Nonexchangeable data require explicit qualification of conformal coverage statements \cite{barber2023}. Consequently, a reference-regime coverage result is not a guarantee for every organization, capability, or missingness mechanism.

For eventual field interventions, panel methods offer more defensible estimands than an unqualified before/after difference. Group-time treatment effects \cite{callaway2021}, heterogeneous event studies \cite{sun2021}, robust imputation estimators \cite{borusyak2024}, and sensitivity to parallel-trend violations \cite{rambachan2023} address distinct design problems. Observational policy learning \cite{athey2021} requires appropriate identification and overlap, while augmented synthetic controls \cite{benmichael2021} impose assumptions about untreated outcomes. We implement none of these as a field causal analysis. Our contrast uses randomized synthetic assignments and a known persistence conversion. This allows a transparent optimizer test while keeping observational identification as future work.

The refined gap is thus an inspectable uncertainty-aware integration and evaluation boundary, rather than the invention of continuous assessment, Bayesian filtering, or causal optimization. The contribution bullets in Section~\ref{sec:introduction} correspond to the measurement implementation, its paired stress protocol, and its constrained decision/data workflow. The audit does not establish that their combination alone meets the novelty threshold of a Transactions paper; that judgment requires stronger field evidence and editorial review.

\section{System Model and Problem Formulation}\label{sec:model}
\subsection{Entities, coordinates, and assumptions}
\begin{figure}[t]
\centering
\begin{tikzpicture}[node distance=4mm and 5mm,box/.style={draw,rounded corners=1pt,align=center,font=\scriptsize,text width=2.9cm,minimum height=7mm},arr/.style={-{Stealth[length=1.5mm]},thin}]
\node[box] (telemetry) {Telemetry and source provenance};
\node[box,right=of telemetry] (survey) {Aggregate survey evidence};
\node[box,below=of telemetry] (cal) {Independent calibration\\and fixed mappings};
\node[box,below=of survey] (mask) {Availability mask\\and evidence graph};
\node[box,below=of cal] (filter) {Augmented Bayesian filter\\capability and survey offset};
\node[box,right=of filter] (audit) {Conditional intervals\\disagreement diagnostics};
\node[box,below=of filter] (predict) {Separate observation\\and outcome forecasts};
\node[box,right=of predict] (policy) {Randomized synthetic effects\\and budgeted screening};
\draw[arr] (telemetry)--(cal);\draw[arr] (survey)--(mask);
\draw[arr] (telemetry.east)--(mask.north west);
\draw[arr] (cal)--(filter);\draw[arr] (mask.south west)--(filter.north east);
\draw[arr] (filter)--(audit);\draw[arr] (filter)--(predict);\draw[arr] (audit)--(policy);
\draw[arr] (predict)--(policy);
\draw[arr,dashed] (filter.west) -- ++(-.25,0) |- (cal.west);
\node[box,below=of predict,text width=6.3cm,xshift=1.7cm] (pilot) {Historical CI pilot: licensed records $\rightarrow$ availability-safe features $\rightarrow$ separate build predictors};
\end{tikzpicture}
\caption{Logical evidence-fusion pipeline. Dashed feedback is sequential updating. Calibration precedes evaluation. The real CI pilot exercises a separate historical feature/prediction path.}\label{fig:architecture}
\end{figure}
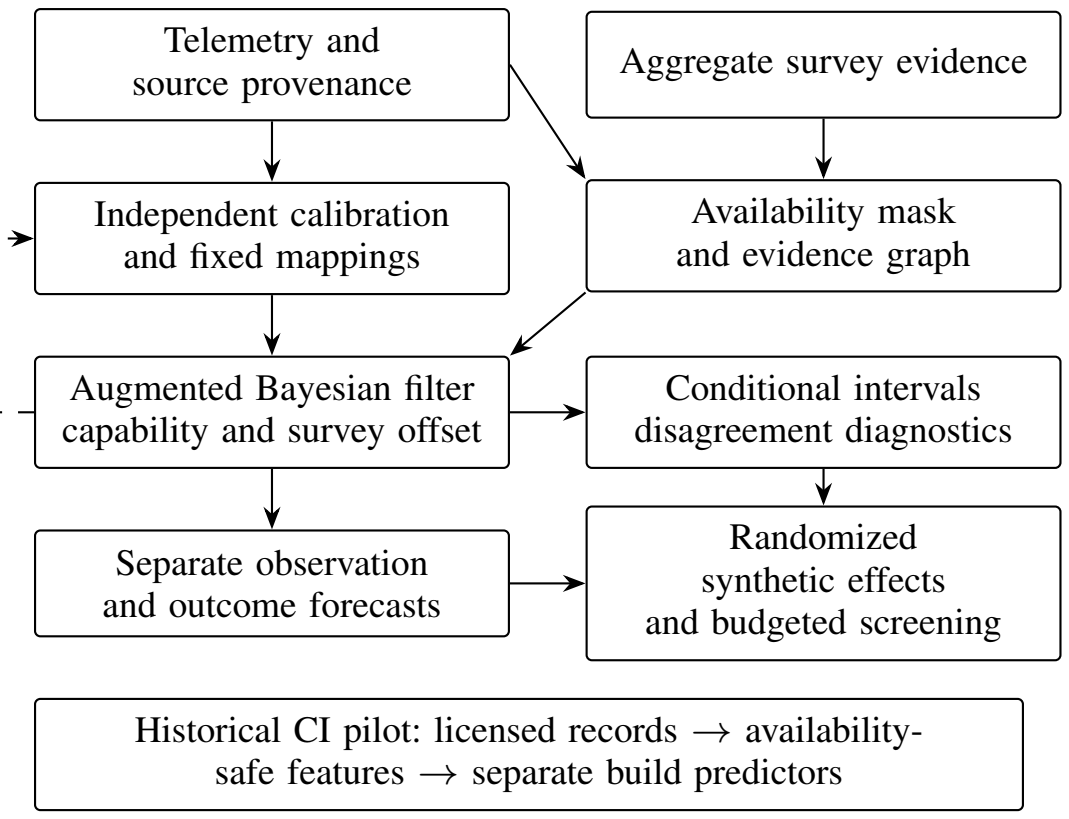

The entities are an organization, week, capability, observation source, and candidate action; Table~\ref{tab:notation} defines the notation. An evidence record contains its source identifier, organization, observation and availability times, capability mapping, value, transformation version, and provenance. The released numeric graph example represents edges from observations to source and capability nodes; it is an interchange schema rather than a claimed graph-database deployment. Figure~\ref{fig:architecture} shows the logical pipeline. No personal identity is required for the model.

Let $o$ identify an organization, $t$ a week, and $k$ one of three technical proxies: delivery, recovery, and security. The labels define simulation coordinates, not empirically validated scales or discrete maturity levels. Fixed unit loadings and an agreed sign anchor each coordinate. With freely estimated loadings, an invertible transformation of the latent states and its inverse on the loadings leave observations unchanged. Persistent survey offsets would likewise be confounded with capability levels without the telemetry anchor. Independent instrument calibration and mapping validation are therefore prerequisites for a field implementation.

We use an exchangeable Gaussian prior within each organization, not a learned multilevel organization hierarchy. The fitted transition is a random walk with a slowly changing survey offset:
\begin{equation}
 \begin{aligned}
 u_{ot}&=u_{o,t-1}+\eta_{ot},\quad u_{ot}=(x_{ot}^{\mathsf T},b_{ot}^{\mathsf T})^{\mathsf T},\\
 \eta_{ot}&\sim\mathcal N(0,Q_u),\quad Q_u=\diag(qI_3,10^{-5}I_3).
 \end{aligned}
 \label{eq:transition}
\end{equation}
Here $x_{ot}\in\R^3$ is capability and $b_{ot}\in\R^3$ is an aggregate survey offset. The initial covariance is $\diag(I_3,0.25I_3)$ and initial mean zero. These are declared assumptions, not priors elicited from SaaS experts. The simulator deliberately adds weak mean reversion and correlated innovations, so the fitted transition is not its exact latent law.

\begin{table}[t]
\caption{Notation and units. Capability units are synthetic and dimensionless.}\label{tab:notation}
\centering\small
\begin{tabular}{ll}
\toprule
Symbol & Meaning\\\midrule
$o,t,k$ & Organization, week, capability\\
$x_{ot},b_{ot}$ & Capability and survey offset in $\R^3$\\
$u_{ot}$ & Augmented state in $\R^6$\\
$z_{ot},H$ & Observations in $\R^9$ and fixed loading matrix\\
$R,R_T$ & Full observation and telemetry covariance\\
$q,Q_u$ & Process variance and augmented covariance\\
$\mathcal O_{ot}$ & Available observation indices\\
$m^-,P^-$ & Prior state mean and covariance\\
$m,P$ & Posterior state mean and covariance\\
$S,K$ & Innovation covariance and Kalman gain\\
$V$ & Preintervention validation observations\\
$\theta_a$ & Instantaneous capability shock of action $a$\\
$\hat p,\beta$ & Calibrated synthetic incident risk and coefficients\\
$v_{oa},c_a$ & Expected risk reduction and action cost\\
$B,z_{oa}$ & Budget and binary action allocation\\
\bottomrule
\end{tabular}
\end{table}

\subsection{Observation and missing-source model}
Each capability has two transformed telemetry sources. At survey weeks, an aggregate continuous score supplies a third channel. The stacked observation model is
\begin{equation}
 z_{ot}=H u_{ot}+\epsilon_{ot},\quad
 H=\begin{bmatrix}A&0\\I_3&I_3\end{bmatrix},\quad
 \epsilon_{ot}\sim\mathcal N(0,R),
 \label{eq:observation}
\end{equation}
where $A\in\R^{6\times3}$ repeats each coordinate twice. This is a Gaussian model of an aggregate score. No ordinal questionnaire likelihood or dynamic item-response model has been implemented.

The two-source covariance for capability $k$ illustrates the dependence retained by the estimator:
\begin{equation}
 R_{T,k}=\begin{bmatrix}
 \sigma_1^2&\rho\sigma_1\sigma_2\\
 \rho\sigma_1\sigma_2&\sigma_2^2
 \end{bmatrix},\qquad R=\diag(R_T,\sigma_s^2 I_3).
 \label{eq:covariance}
\end{equation}
The implementation estimates the full $6\times6$ telemetry covariance from separate synthetic calibration draws; the displayed blocks describe their generating law. The reference standard deviations are 0.30 and 0.40, $\rho=\SourceCorrelation{}$, and survey noise standard deviation is 0.20. Sharing evidence therefore does not create two independent measurements. Survey offsets are generated with mean 0.50 and standard deviation 0.25, while the estimator initially centers their prior at zero.

Let $D_t=z_{o,1:t}^{\mathcal O}$ contain only available observations, with fixed $H,R,q$. Define $\ell_t(u)$ as their Gaussian likelihood using $H_{\mathcal O}$ and $R_{\mathcal O\mathcal O}$. Conditional Bayesian inference follows
\begin{equation}
 \begin{aligned}
 p(u_{ot}\mid D_t)&\propto \ell_t(u_{ot})p(u_{ot}\mid D_{t-1}),\\
 p(u_{ot}\mid D_{t-1})&=\int p(u_{ot}\mid v)p(v\mid D_{t-1})dv.
 \end{aligned}
 \label{eq:posterior}
\end{equation}
The proportionality normalizes over $u_{ot}$. Selecting rows of $H$ and the corresponding principal submatrix of $R$ preserves source dependence when observations disappear. An empty index set gives a prediction-only step. This likelihood ignores the missingness process. It is appropriate for the imposed random-loss mechanism, but is not an identified correction for informative loss \cite{rubin1976,miao2016}.

The measurement objective is accurate state recovery with calibrated conditional intervals, while the operational objectives are next-week observed-telemetry prediction and synthetic incident risk prediction. These objectives are evaluated separately. A third objective selects at most one action per organization under a common budget. The decision model assumes a transportable action shock in the simulator; actual business maturity, tenant isolation, customer impact, cloud cost, and organizational culture are outside the measured scope.

\section{Filtering and Intervention Screening}\label{sec:method}
\subsection{Sequential measurement update}
For each organization, the transition in~\eqref{eq:transition} predicts the augmented state using only prior information:
\begin{equation}
 m^-_{ot}=m_{o,t-1},\qquad P^-_{ot}=P_{o,t-1}+Q_u.
 \label{eq:prediction}
\end{equation}
For the available indices $\mathcal O=\mathcal O_{ot}$, the innovation covariance and gain are
\begin{equation}
 \begin{aligned}
 S_{ot}&=H_{\mathcal O}P^-_{ot}H_{\mathcal O}^{\mathsf T}+R_{\mathcal O\mathcal O},\\
 K_{ot}&=P^-_{ot}H_{\mathcal O}^{\mathsf T}S_{ot}^{-1}.
 \end{aligned}
 \label{eq:gain}
\end{equation}
The code uses a linear solve rather than explicitly inverting $S$. The update uses the Joseph covariance form to retain positive semidefiniteness numerically:
\begin{equation}
 \begin{aligned}
 m_{ot}&=m^-_{ot}+K_{ot}(z_{ot}^{\mathcal O}-H_{\mathcal O}m^-_{ot}),\\
 P_{ot}&=(I-K_{ot}H_{\mathcal O})P^-_{ot}(I-K_{ot}H_{\mathcal O})^{\mathsf T}\\
             &\quad+K_{ot}R_{\mathcal O\mathcal O}K_{ot}^{\mathsf T}.
 \end{aligned}
 \label{eq:update}
\end{equation}
An offset can absorb persistent survey disagreement while repeated telemetry anchors the capability component. It cannot distinguish a newly biased telemetry source from an actual capability change without additional assumptions. Similarly, if only surveys are observed for a long period, the prior and earlier anchors determine how evidence is divided between capability and offset. Wider uncertainty is an appropriate result, not a failed score that should be hidden.

We choose $q$ from a fixed grid using preintervention validation observations. Write the prediction residual as $r_{otk}(q)=y_{otk}-\hat y_{otk}(q)$ and let $\mathcal Q=\{.0016,.0049,.0144\}$:
\begin{equation}
 \hat q=\argmin_{q\in\mathcal Q}\frac{1}{2|V|}\sum_{(o,t,k)\in V}
 \left[\log(2\pi v_{otk}(q))+\frac{r_{otk}(q)^2}{v_{otk}(q)}\right].
 \label{eq:tuning}
\end{equation}
Here $V$ includes available first-source readings from validation organizations in weeks 32--63, $\hat y$ is a one-week-ahead prediction, and $v$ includes state, process, and measurement variance. Test latent truth never enters this choice. EWMA receives a parallel validation search over its three prescribed smoothing factors. Other ablations share the chosen $q$, making them mechanism removals rather than independently optimized competitors.

For coordinate $k$, the conditional 90\% credible interval is
\begin{equation}
 C^x_{otk}=\left[m_{otk}-1.64485\sqrt{P_{ot,kk}},
                    m_{otk}+1.64485\sqrt{P_{ot,kk}}\right].
 \label{eq:credible}
\end{equation}
This interval conditions on fitted hyperparameters and fixed mappings; it does not integrate uncertainty in calibration, capability meaning, or logistic coefficients. Under the fitted random walk, an available next-week telemetry reading instead has predictive variance
\begin{equation}
 y_{o,t+1,k}^{(1)}\mid z_{o,1:t}^{\mathcal O}
 \sim\mathcal N(m_{otk},\;P_{ot,kk}+\hat q+R_{T,2k-1,2k-1}).
 \label{eq:telemetryforecast}
\end{equation}
The distinction between~\eqref{eq:credible} and~\eqref{eq:telemetryforecast} is essential: covering a noisy observation does not demonstrate that the latent coordinate is calibrated.

\begin{algorithm}{Source-aware sequential evidence update}\label{alg:filter}
\begin{algorithmic}[1]
\Require Time-ordered available observations, $H,R,Q_u$, prior $(m_0,P_0)$
\Ensure Capability means and conditional covariance at every week
\State Initialize $m\gets m_0$, $P\gets P_0$
\For{each week $t$ in ascending availability order}
\State Predict $(m^-,P^-)$ using~\eqref{eq:prediction}
\State Select $\mathcal O_t$, $H_{\mathcal O_t}$, and $R_{\mathcal O_t\mathcal O_t}$
\If{$\mathcal O_t$ is empty}
\State $(m,P)\gets(m^-,P^-)$
\Else
\State Solve for $K$ with~\eqref{eq:gain}; update with~\eqref{eq:update}
\EndIf
\State Save the first three coordinates and their covariance
\EndFor
\State \Return the saved trajectory and provenance identifiers
\end{algorithmic}
\end{algorithm}

Algorithm~\ref{alg:filter} terminates at the last observed week. With state dimension $d=6$ and at most $r=9$ channels, dense updates cost $O(d^2r+dr^2+r^3)$ per week and working state storage costs $O(d^2+r^2)$. Organizations can be processed independently, so cost is linear in their number at fixed $d,r,T$. Saving all trajectories adds $O(NTd^2)$ storage. The implementation's returned arrays retain complete trajectories, so measured memory is not a claim of constant total storage.

\subsection{Randomized synthetic effects and action utility}
The simulator assigns each training organization to a control arm or one of three capability-specific shocks at week 64, with six organizations per arm. Let $g_o$ be the available-source mean in weeks 64--79 minus its mean in weeks 48--63. For action $a$, the shock estimator is
\begin{equation}
 \hat\theta_a=
 \frac{\overline g_{A=a}-\overline g_{A=0}}
 {16^{-1}\sum_{j=0}^{15}\phi^j},\qquad \phi=.99.
 \label{eq:effect}
\end{equation}
The denominator converts a window-average contrast to an instantaneous shock using known simulator persistence. This is a privileged structural assumption, not an inferred real-world intervention law. Organizations are bootstrapped within arms to obtain effect draws. The randomized, randomly missing reference setting supports this contrast; applying the same expression after endogenous field adoption would not identify a causal effect.

At week 88, a separate logistic risk model, trained on contemporaneous preintervention training states and incident labels, supplies $\hat p(x)=\operatorname{logit}^{-1}(\beta_0+\beta^{\mathsf T}x)$. It differs from the next-week predictor used for forecast evaluation. For state draws $x^{(j)}$ from the conditional Gaussian filter and bootstrap effect draws $\theta_a^{(j)}$, expected instantaneous utility is approximated by
\begin{equation}
 \hat v_{oa}=\frac{1}{M}\sum_{j=1}^{M}
 \left[\hat p(x_o^{(j)})-\hat p(x_o^{(j)}+\theta_a^{(j)})\right],
 \qquad M=256.
 \label{eq:utility}
\end{equation}
The conservative policy substitutes the tenth percentile of these risk reductions. It is a hybrid conditional-state/bootstrap sensitivity rule, not a fully Bayesian posterior over all causal and prediction parameters. The fitted logistic coefficients are held fixed.

With integer cost units $c=(5,8,4)$ and budget $B=48$, selection solves
\begin{equation}
 \begin{aligned}
 \max_{z_{oa}\in\{0,1\}}\quad&\sum_{o,a}v_{oa}z_{oa},\\
 \text{subject to}\quad&\sum_a z_{oa}\le1\quad\forall o,\qquad
 \sum_{o,a}c_a z_{oa}\le B.
 \end{aligned}
 \label{eq:budget}
\end{equation}
Deferral has zero cost and zero utility, so negative utility never forces an action. These costs are synthetic decision units, not measured cloud or labor expenditure. The oracle solves the same problem using instantaneous true simulator risk reductions at the same week. Regret is oracle benefit minus achieved true benefit; it is observable only because the outcome law is known.

\begin{algorithm}{Budgeted action screening}\label{alg:policy}
\begin{algorithmic}[1]
\Require State draws, randomized effect draws, risk model, costs $c_a$, budget $B$
\Ensure At most one selected action per organization
\State Compute mean or conservative utilities from~\eqref{eq:utility}
\State Initialize $D(0,0)=0$; all unreachable budget states to $-\infty$
\For{organization $o=1,\ldots,N$}
\For{reachable spent budget $b=0,\ldots,B$}
\For{candidate $a$ including zero-cost deferral}
\If{$b+c_a\le B$}
\State Update $D(o,b+c_a)$ from $D(o-1,b)+v_{oa}$; retain predecessor
\EndIf
\EndFor
\EndFor
\EndFor
\State Backtrack the highest-valued reachable final state
\State \Return chosen actions and saved uncertainty/evidence
\end{algorithmic}
\end{algorithm}

Algorithm~\ref{alg:policy} is exact for the discrete multiple-choice budget in~\eqref{eq:budget}. Its dynamic program costs $O(NAB)$ time and $O(NB)$ storage with backtracking, in addition to risk prediction for the draws. A finer cost discretization increases $B$. Interference between organizations, simultaneous actions, delayed benefits, and intervention-induced measurement changes are not modeled.

\section{Experimental Protocol}\label{sec:experiments}
\subsection{Paired synthetic cohorts}
Ten seeds each generate \OrgCount{} organizations over \Weeks{} weeks. Training, validation, and test groups contain \TrainOrgs{}, \ValOrgs{}, and \TestOrgs{} organizations, respectively. Each seed defines one latent world and independent instrument-calibration draws. Four observation regimes share that latent world: reference loss probability 0.15; extensive loss probability 0.60; informative loss probability $0.10+0.65\operatorname{logit}^{-1}(-2x)$; and a 1.8-fold measurement-noise standard-deviation increase from week 80. Each organization also has a six-week outage in one telemetry source. Survey observations occur every fourth week with additional independent loss probability 0.10. These are designed stressors rather than estimates of real collection failure rates. Table~\ref{tab:configuration} records the configuration.

The generator has persistence 0.99, process standard deviation 0.07, and cross-capability innovation correlation 0.25. Capability shocks are 0.35, 0.45, and 0.25 in the three coordinates. A synthetic incident indicator follows a specified logistic law of the latent state plus a small periodic term. It is not a production incident dataset. All test metrics use weeks 64--103. Validation uses only observations in weeks 32--63, before any shock. Outcome calibration uses preintervention training labels. The latent truth is reserved for evaluation and oracle calculation.

\begin{table}[t]
\caption{Prespecified simulation configuration. These are assumptions, not empirical SaaS estimates.}\label{tab:configuration}
\centering\small\setlength{\tabcolsep}{4pt}
\begin{tabular}{ll}\toprule
Setting & Value\\\midrule
Seeds / observation worlds & \Seeds{} / \Worlds{}\\
Organizations, weeks per world & \OrgCount{}, \Weeks{}\\
Training / validation / test organizations & \TrainOrgs{} / \ValOrgs{} / \TestOrgs{}\\
Telemetry channels / survey channels & 6 / 3\\
Calibration noise draws per seed & 1,200\\
Survey period / synthetic intervention week & 4 weeks / 64\\
State prior / bias prior variance & 1.00 / 0.25\\
Bias transition variance & $10^{-5}$\\
Validation process variances & .0016, .0049, .0144\\
EWMA smoothing candidates & .2, .5, .8\\
Conformal window / target miscoverage & 32 / .10\\
ACI step / PID scale, saturation & .005 / .10, 10\\
Action cost units / common budget & (5,8,4) / 48\\
Effect bootstrap / decision draws & 400 / 256\\
\bottomrule\end{tabular}
\end{table}

\subsection{Comparators and task separation}
State comparators are the current available-source telemetry mean with carry-forward, survey-only carry-forward, tuned EWMA, and uncorrected Gaussian fusion without a survey-offset state. Ablations remove surveys, discard source covariance, or reset the capability prior each week while retaining the offset state. These are eight directly executed estimators. They are conventional baselines and mechanism tests, not claimed reproductions of the questionnaire instrument, CMAM, or BayesInsights.

Observed-telemetry intervals use the same one-step Twin point forecast. Gaussian intervals follow~\eqref{eq:telemetryforecast}; rolling conformal intervals use the exact order statistic of the last 32 available absolute errors with rank $\lceil(n+1)(1-\alpha)\rceil$. ACI implements the original unprojected recurrence $\alpha_{t+1}=\alpha_t+0.005(\alpha-\mathrm{err}_t)$ \cite{aci2021}. Extended quantiles produce an unbounded or empty set when required, rather than silently clipping the nominal level. PID adapts the paper's Eq.~5 \cite{pid2023}: a rolling-quantile scorecaster plus a tangent integral of accumulated miscoverage, with scale 0.10 and saturation constant 10. It does not reproduce the original Theta scorecaster or benchmark. Both controller settings are prespecified. Updates use only observed labels after issuing their intervals; missing targets do not trigger feedback. We do not claim that these are the two strongest current SaaS assessment methods.

\subsection{Historical real-data pilot}
The TravisTorrent Figshare deposit is licensed CC BY 4.0; its 2016-03-05 snapshot was downloaded and checksum-verified \cite{beller2017}. The source has \CIRawJobs{} job rows. We deduplicate build identifiers, exclude unsupported statuses and invalid durations, and retain projects with at least 40 builds. This yields \CIEligibleBuilds{} eligible builds before feature-history and split exclusions. The nine features use previously completed build outcomes, trailing failure rates, earlier log durations, elapsed time, history count, and target weekday. A build becomes eligible as historical evidence only after its reported start plus duration and a 24-hour embargo. This completion time is a metadata proxy, not an independently verified finish event.

Projects are assigned by a deterministic hash, with no identity overlap between training, validation, and test. Calendar cutoffs further require training targets to precede validation and validation to precede testing. Training and validation labels are purged when their completion proxy falls within 24 hours of the next cutoff, preventing boundary-label leakage. Held-out projects retain their own previously available histories, which is appropriate for online prediction but is not a cold-history experiment. Fork and clone independence are unverified. Table~\ref{tab:ci-data} reports the resulting cohorts. Training is capped at \CIUsedTrainRows{} rows using one fixed sample.

\begin{table}[t]
\caption{Historical CI cohorts after availability purges. Projects are distinct between partitions. The failure label includes failed and errored builds, not production change failures.}\label{tab:ci-data}
\centering\small\begin{tabular}{lrrr}
\toprule
Partition & Builds & Projects & Failure rate \\
\midrule
Train & 129,757 & 662 & 31.1\% \\
Validation & 11,153 & 211 & 25.0\% \\
Test & 13,074 & 224 & 19.5\% \\
\bottomrule
\end{tabular}

\end{table}

We fit logistic regression, a 100-tree random forest, and histogram gradient boosting, alongside training prevalence and trailing failure rate. Boosting chooses between 60 and 100 iterations using validation Brier score. Other hyperparameters are fixed; no comprehensive state-of-the-art search is claimed. Five algorithm seeds use the same split and training sample, so their variation is not independent population evidence. Project-bootstrap intervals use first-seed held-out predictions and resample whole projects.

\subsection{Metrics, inference, and execution}
Latent RMSE, 90\% state coverage, mean width, and interval score evaluate measurement. Next-week incident Brier score and AUROC evaluate a separate logistic calibration fitted on each method's estimated training states. Telemetry coverage, width, and central interval score evaluate observation forecasts; the score is width plus $2/\alpha$ times each missed-tail distance \cite{gneiting2007,bracher2021}. Empty sets receive infinite interval score and zero length, while unbounded sets retain infinite length and score. Such sets must not be removed to improve a mean.

We summarize simulation metrics across independent seeds, not individual weeks. Paired seed-level RMSE differences use a fully enumerated sign-flip test with Holm correction across seven comparisons. Its inferential null requires sign exchangeability or symmetry of paired differences; enumerating signs does not remove that assumption. Bootstrap intervals resample independent seeds for these contrasts. They quantify simulator variability, not sampling uncertainty across real organizations. Runtime and memory come from the executed CPU environment, recorded package versions, process peak resident set size, and a separate allocation-traced scaling run. Python allocation tracing excludes native-library allocations. The implementation uses NumPy, SciPy, pandas, and scikit-learn; no accelerator or unobserved processor model is claimed.

\section{Results and Discussion}\label{sec:results}
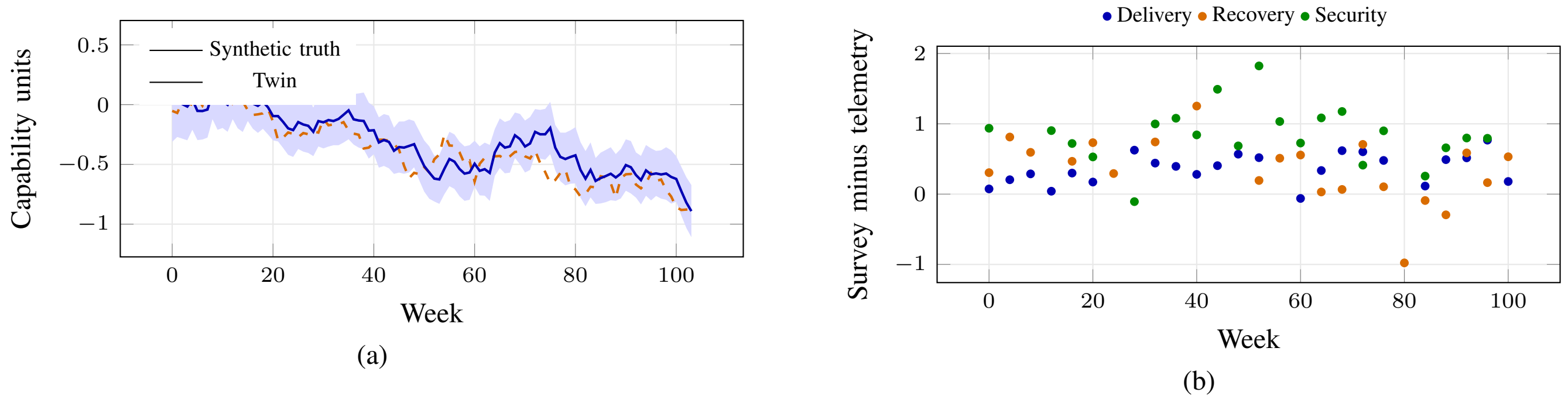
\begin{figure*}[t]
\centering
\begin{minipage}{.485\textwidth}\centering\begin{tikzpicture}
\begin{axis}[width=.98\linewidth,height=4.25cm,
tick label style={font=\scriptsize},label style={font=\small},legend style={font=\scriptsize,draw=none},
grid=major,grid style={gray!18},unbounded coords=jump,xlabel={Week},ylabel={Capability units},legend pos=north west]
\addplot[name path=L,draw=none] table[x=week,y=lower,col sep=comma] {figures/data/panel01.csv};
\addplot[name path=U,draw=none] table[x=week,y=upper,col sep=comma] {figures/data/panel01.csv};
\addplot[blue!15,forget plot] fill between[of=L and U];
\addplot[orange!85!black,thick,dashed] table[x=week,y=truth,col sep=comma] {figures/data/panel01.csv};
\addlegendentry{Synthetic truth}
\addplot[blue!70!black,thick] table[x=week,y=mean,col sep=comma] {figures/data/panel01.csv};
\addlegendentry{Twin}

\end{axis}
\end{tikzpicture}\par\small (a)\end{minipage}\hfill
\begin{minipage}{.485\textwidth}\centering\begin{tikzpicture}
\begin{axis}[width=.98\linewidth,height=4.25cm,
tick label style={font=\scriptsize},label style={font=\small},legend style={font=\scriptsize,draw=none},
grid=major,grid style={gray!18},unbounded coords=jump,xlabel={Week},ylabel={Survey minus telemetry},legend columns=3,legend style={at={(.5,1.03)},anchor=south}]
\addplot[blue!70!black,only marks,mark=*,mark size=1.2pt] table[x=week,y=gap0,col sep=comma] {figures/data/panel02.csv};
\addlegendentry{Delivery}
\addplot[orange!85!black,only marks,mark=*,mark size=1.2pt] table[x=week,y=gap1,col sep=comma] {figures/data/panel02.csv};
\addlegendentry{Recovery}
\addplot[green!55!black,only marks,mark=*,mark size=1.2pt] table[x=week,y=gap2,col sep=comma] {figures/data/panel02.csv};
\addlegendentry{Security}

\end{axis}
\end{tikzpicture}\par\small (b)\end{minipage}
\caption{Evidence trajectories in one synthetic held-out organization. (a) Delivery state and conditional 90\% credible band. (b) Raw survey--telemetry disagreement at available survey times.}\label{fig:evidence}
\end{figure*}
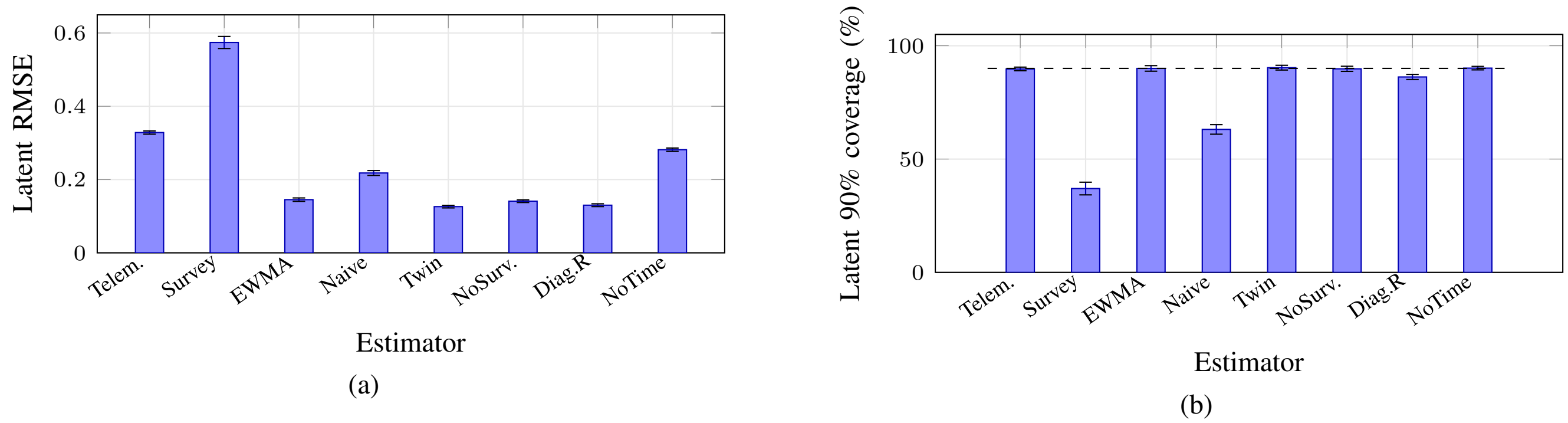
\begin{figure*}[t]
\centering
\begin{minipage}{.485\textwidth}\centering\begin{tikzpicture}
\begin{axis}[width=.98\linewidth,height=4.25cm,
tick label style={font=\scriptsize},label style={font=\small},legend style={font=\scriptsize,draw=none},
grid=major,grid style={gray!18},unbounded coords=jump,xlabel={Estimator},ylabel={Latent RMSE},xtick={1,2,3,4,5,6,7,8},xticklabels={Telem.,Survey,EWMA,Naive,Twin,NoSurv.,Diag.R,NoTime},x tick label style={rotate=35,anchor=east},ymin=0]
\addplot[ybar,fill=blue!45,draw=blue!70!black,bar width=9pt,error bars/.cd,y dir=both,y explicit] table[x=method,y=rmse,y error=rmse_ci,col sep=comma] {figures/data/panel03.csv};

\end{axis}
\end{tikzpicture}\par\small (a)\end{minipage}\hfill
\begin{minipage}{.485\textwidth}\centering\begin{tikzpicture}
\begin{axis}[width=.98\linewidth,height=4.25cm,
tick label style={font=\scriptsize},label style={font=\small},legend style={font=\scriptsize,draw=none},
grid=major,grid style={gray!18},unbounded coords=jump,xlabel={Estimator},ylabel={Latent 90\% coverage (\%)},xtick={1,2,3,4,5,6,7,8},xticklabels={Telem.,Survey,EWMA,Naive,Twin,NoSurv.,Diag.R,NoTime},x tick label style={rotate=35,anchor=east},ymin=0,ymax=105]
\addplot[ybar,fill=blue!45,draw=blue!70!black,bar width=9pt,error bars/.cd,y dir=both,y explicit] table[x=method,y=coverage,y error=coverage_ci,col sep=comma] {figures/data/panel04.csv};
\addplot[black,dashed,domain=.5:8.5] {90};

\end{axis}
\end{tikzpicture}\par\small (b)\end{minipage}
\caption{Reference-regime latent-state estimates across independent synthetic cohorts. (a) RMSE (lower is better). (b) Credible coverage; the dashed line is the 90\% target. Error bars summarize uncertainty across seeds, not organizations.}\label{fig:state}
\end{figure*}
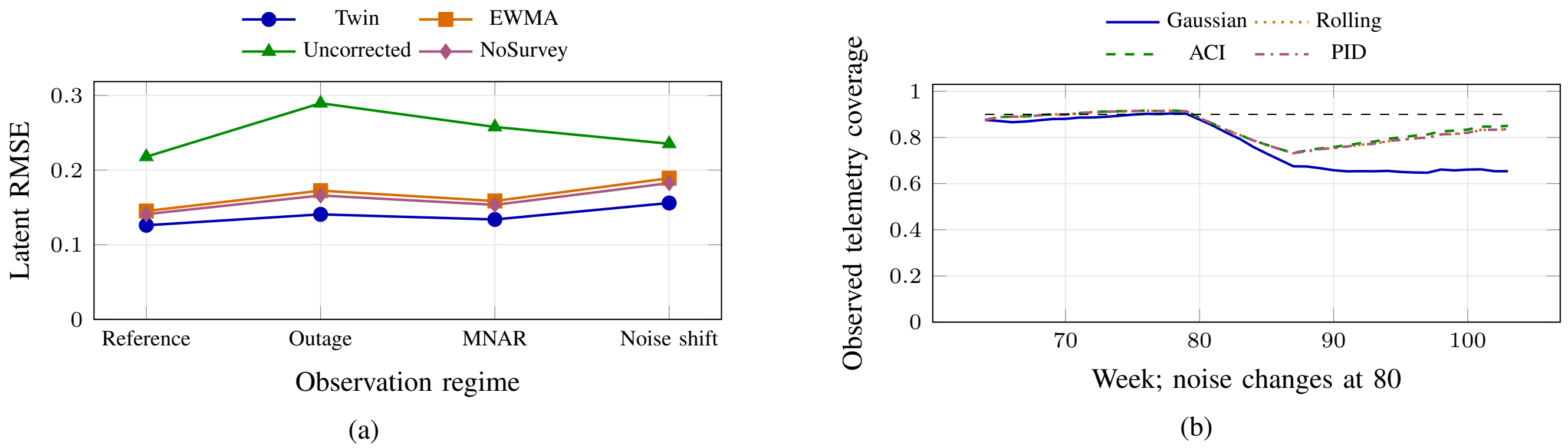
\begin{figure*}[t]
\centering
\begin{minipage}{.485\textwidth}\centering\begin{tikzpicture}
\begin{axis}[width=.98\linewidth,height=4.25cm,
tick label style={font=\scriptsize},label style={font=\small},legend style={font=\scriptsize,draw=none},
grid=major,grid style={gray!18},unbounded coords=jump,xlabel={Observation regime},ylabel={Latent RMSE},xtick={1,2,3,4},xticklabels={Reference,Outage,MNAR,Noise shift},ymin=0,legend columns=2,legend style={at={(.5,1.03)},anchor=south}]
\addplot[blue!70!black,thick,mark=*] table[x=scenario,y=Twin,col sep=comma] {figures/data/panel05.csv};
\addlegendentry{Twin}
\addplot[orange!85!black,thick,mark=square*] table[x=scenario,y=EWMA,col sep=comma] {figures/data/panel05.csv};
\addlegendentry{EWMA}
\addplot[green!55!black,thick,mark=triangle*] table[x=scenario,y=Uncorrected,col sep=comma] {figures/data/panel05.csv};
\addlegendentry{Uncorrected}
\addplot[magenta!65!black,thick,mark=diamond*] table[x=scenario,y=NoSurvey,col sep=comma] {figures/data/panel05.csv};
\addlegendentry{NoSurvey}

\end{axis}
\end{tikzpicture}\par\small (a)\end{minipage}\hfill
\begin{minipage}{.485\textwidth}\centering\begin{tikzpicture}
\begin{axis}[width=.98\linewidth,height=4.25cm,
tick label style={font=\scriptsize},label style={font=\small},legend style={font=\scriptsize,draw=none},
grid=major,grid style={gray!18},unbounded coords=jump,xlabel={Week; noise changes at 80},ylabel={Observed telemetry coverage},ymin=0,ymax=1.03,legend columns=2,legend style={at={(.5,1.03)},anchor=south}]
\addplot[blue!70!black,thick] table[x=week,y=Gaussian,col sep=comma] {figures/data/panel06.csv};
\addlegendentry{Gaussian}
\addplot[orange!85!black,thick,dotted] table[x=week,y=Rolling,col sep=comma] {figures/data/panel06.csv};
\addlegendentry{Rolling}
\addplot[green!55!black,thick,dashed] table[x=week,y=ACI,col sep=comma] {figures/data/panel06.csv};
\addlegendentry{ACI}
\addplot[magenta!65!black,thick,dashdotted] table[x=week,y=PID,col sep=comma] {figures/data/panel06.csv};
\addlegendentry{PID}
\addplot[black,dashed,domain=64:103] {.9};

\end{axis}
\end{tikzpicture}\par\small (b)\end{minipage}
\caption{Observation stress tests. (a) Latent-state recovery across four paired regimes. (b) Eight-week trailing observed-telemetry coverage after the week-80 noise increase. ACI and PID target future observations, not latent states.}\label{fig:robustness}
\end{figure*}
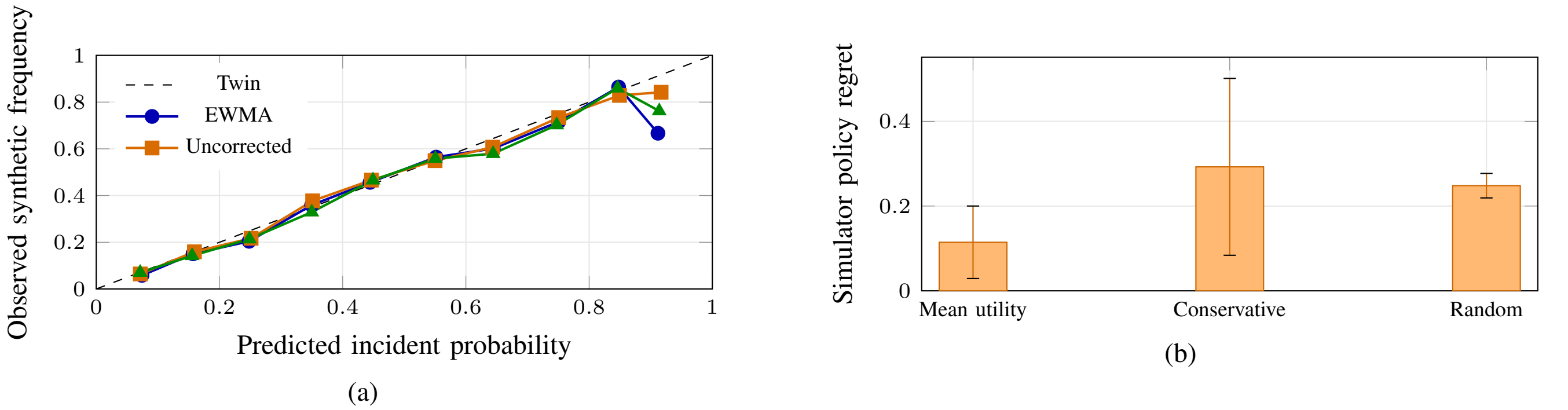
\begin{figure*}[t]
\centering
\begin{minipage}{.485\textwidth}\centering\begin{tikzpicture}
\begin{axis}[width=.98\linewidth,height=4.25cm,
tick label style={font=\scriptsize},label style={font=\small},legend style={font=\scriptsize,draw=none},
grid=major,grid style={gray!18},unbounded coords=jump,xlabel={Predicted incident probability},ylabel={Observed synthetic frequency},xmin=0,xmax=1,ymin=0,ymax=1,legend pos=north west]
\addplot[black,dashed,domain=0:1] {x};
\addplot[blue!70!black,thick,mark=*] table[x=Twinx,y=Twiny,col sep=comma] {figures/data/panel07.csv};
\addlegendentry{Twin}
\addplot[orange!85!black,thick,mark=square*] table[x=EWMAx,y=EWMAy,col sep=comma] {figures/data/panel07.csv};
\addlegendentry{EWMA}
\addplot[green!55!black,thick,mark=triangle*] table[x=Uncorrectedx,y=Uncorrectedy,col sep=comma] {figures/data/panel07.csv};
\addlegendentry{Uncorrected}

\end{axis}
\end{tikzpicture}\par\small (a)\end{minipage}\hfill
\begin{minipage}{.485\textwidth}\centering\begin{tikzpicture}
\begin{axis}[width=.98\linewidth,height=4.25cm,
tick label style={font=\scriptsize},label style={font=\small},legend style={font=\scriptsize,draw=none},
grid=major,grid style={gray!18},unbounded coords=jump,ylabel={Simulator policy regret},xtick={1,2,3},xticklabels={Mean utility,Conservative,Random},ymin=0]
\addplot[ybar,fill=orange!55,draw=orange!80!black,bar width=22pt,error bars/.cd,y dir=both,y explicit] table[x=policy,y=regret,y error=ci,col sep=comma] {figures/data/panel08.csv};
\end{axis}
\end{tikzpicture}\par\small (b)\end{minipage}
\caption{Downstream diagnostics. (a) Pooled synthetic incident calibration. (b) Policy regret against a simulator oracle at the same instant and budget; lower is better. Conservative deferral can sacrifice benefit.}\label{fig:decision}
\end{figure*}
\begin{figure*}[t]
\centering
\begin{minipage}{.485\textwidth}\centering\begin{tikzpicture}
\begin{axis}[width=.98\linewidth,height=4.25cm,
tick label style={font=\scriptsize},label style={font=\small},legend style={font=\scriptsize,draw=none},
grid=major,grid style={gray!18},unbounded coords=jump,xlabel={Process variance $q$ ($10^{-3}$)},ylabel={Validation telemetry NLL}]
\addplot[blue!70!black,thick,mark=*,error bars/.cd,y dir=both,y explicit] table[x=q,y=nll,y error=ci,col sep=comma] {figures/data/panel09.csv};
\end{axis}
\end{tikzpicture}\par\small (a)\end{minipage}\hfill
\begin{minipage}{.485\textwidth}\centering\begin{tikzpicture}
\begin{axis}[width=.98\linewidth,height=4.25cm,
tick label style={font=\scriptsize},label style={font=\small},legend style={font=\scriptsize,draw=none},
grid=major,grid style={gray!18},unbounded coords=jump,xlabel={Organizations (104 weeks each)},ylabel={Filter seconds with tracing},xmin=0,ymin=0]
\addplot[blue!70!black,thick,mark=*] table[x=organizations,y=seconds,col sep=comma] {figures/data/panel10.csv};

\end{axis}
\end{tikzpicture}\par\small (b)\end{minipage}
\caption{Sensitivity and computation. (a) Preintervention validation negative log likelihood across process variances. (b) Single CPU scaling run with Python allocation tracing enabled; no timing confidence interval is asserted.}\label{fig:sensitivity}
\end{figure*}
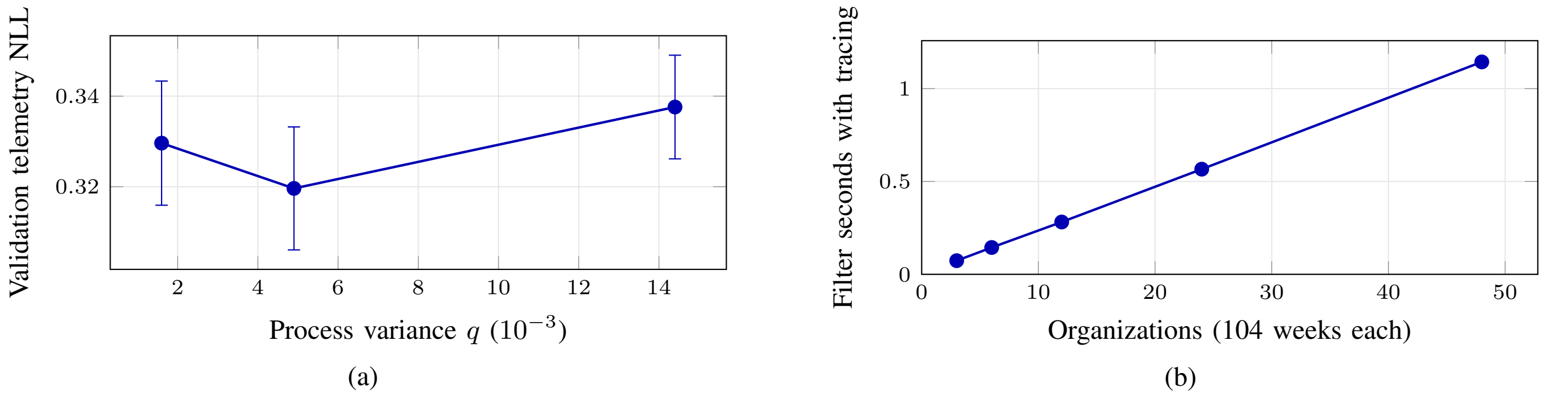
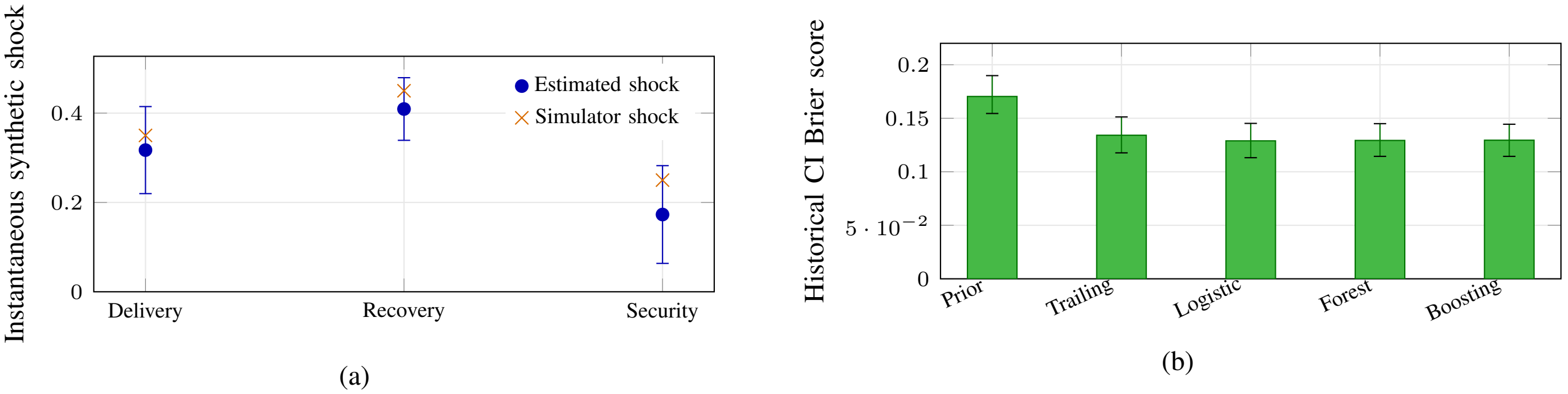
\begin{figure*}[t]
\centering
\begin{minipage}{.485\textwidth}\centering\begin{tikzpicture}
\begin{axis}[width=.98\linewidth,height=4.25cm,
tick label style={font=\scriptsize},label style={font=\small},legend style={font=\scriptsize,draw=none},
grid=major,grid style={gray!18},unbounded coords=jump,ylabel={Instantaneous synthetic shock},xtick={1,2,3},xticklabels={Delivery,Recovery,Security},ymin=0,legend pos=north east]
\addplot[blue!70!black,only marks,mark=*,error bars/.cd,y dir=both,y explicit] table[x=action,y=estimate,y error=ci,col sep=comma] {figures/data/panel11.csv};
\addlegendentry{Estimated shock}
\addplot[orange!85!black,only marks,mark=x,mark size=3pt] table[x=action,y=truth,col sep=comma] {figures/data/panel11.csv};
\addlegendentry{Simulator shock}

\end{axis}
\end{tikzpicture}\par\small (a)\end{minipage}\hfill
\begin{minipage}{.485\textwidth}\centering\begin{tikzpicture}
\begin{axis}[width=.98\linewidth,height=4.25cm,
tick label style={font=\scriptsize},label style={font=\small},legend style={font=\scriptsize,draw=none},
grid=major,grid style={gray!18},unbounded coords=jump,ylabel={Historical CI Brier score},xtick={1,2,3,4,5},xticklabels={Prior,Trailing,Logistic,Forest,Boosting},x tick label style={rotate=25,anchor=east},ymin=0,ymax=.22]
\addplot[ybar,fill=green!45!gray,draw=green!45!black,bar width=16pt,error bars/.cd,y dir=both,y explicit] table[x=method,y=brier,y error minus=minus,y error plus=plus,col sep=comma] {figures/data/panel12.csv};
\end{axis}
\end{tikzpicture}\par\small (b)\end{minipage}
\caption{External scope boundary. (a) Estimated randomized synthetic action shocks, with across-seed confidence bars and simulator truths. (b) Historical OSS CI prediction with project-bootstrap intervals; this panel does not validate SaaS maturity.}\label{fig:pilot}
\end{figure*}

\subsection{Reference measurement and practical outcome gap}
Figure~\ref{fig:evidence} shows one complete held-out synthetic trajectory and the raw evidence disagreement. Survey offsets remain visible while the filtered capability follows the telemetry-supported state. This is a diagnostic example, not a representative case selected to prove performance. Its seed, organization, readings, and posterior are included in the raw files.

Across all reference cohorts, Twin achieves RMSE \TwinRMSE{}, versus \EWMARMSE{} for tuned EWMA, a relative reduction of \RelativeRMSEGain{}\%. The paired mean difference is \PairedDifference{}, with a bootstrap 95\% interval $[\PairedDifferenceLow{},\PairedDifferenceHigh{}]$ and Holm-adjusted sign-flip $p=\PairedHolmP{}$ under the stated null. Table~\ref{tab:main} and Fig.~\ref{fig:state} show that the largest degradation occurs when a biased survey is treated as unbiased or when temporal information is removed. The no-survey ablation has RMSE \NoSurveyRMSE{}; hence surveys can add information after their offset is modeled, even though the naive fusion baseline is worse than telemetry smoothing.

\begin{table*}[t]
\caption{Reference synthetic results across ten seeds. RMSE is mean $\pm$ sample standard deviation; coverage and width are conditional latent intervals, while Brier evaluates next-week synthetic incidents. Bold and underline identify the lowest and second-lowest directly comparable RMSE. Better state recovery need not imply a substantial outcome improvement.}\label{tab:main}
\centering\small\begin{tabular}{lrrrr}
\toprule
Estimator & RMSE $\downarrow$ & Cov. (\%) & Width $\downarrow$ & Brier $\downarrow$ \\
\midrule
Telemetry & 0.3283 $\pm$ 0.0062 & 89.8 & 1.072 & 0.2060 \\
Survey & 0.5741 $\pm$ 0.0230 & 37.0 & 0.730 & 0.2075 \\
EWMA & 0.1452 $\pm$ 0.0065 & 90.0 & 0.475 & 0.2005 \\
Uncorrected fusion & 0.2179 $\pm$ 0.0093 & 63.1 & 0.409 & 0.2016 \\
Twin & \textbf{0.1261 $\pm$ 0.0050} & 90.3 & 0.415 & 0.2003 \\
No survey & 0.1410 $\pm$ 0.0051 & 89.8 & 0.458 & 0.2004 \\
Diagonal covariance & \underline{0.1300 $\pm$ 0.0054} & 86.2 & 0.386 & 0.2004 \\
No dynamics & 0.2814 $\pm$ 0.0064 & 90.1 & 0.926 & 0.2059 \\
\bottomrule
\end{tabular}

\end{table*}

The full model's \TwinCoverage{}\% state coverage is close to its 90\% nominal level in this reference setting. Diagonalizing source covariance narrows the interval and lowers coverage to \DiagonalRCoverage{}\%, consistent with treating overlapping evidence as too informative. Uncorrected fusion covers only \UncorrectedCoverage{}\%. This is not a general proof that survey offsets are constant or that Gaussian intervals are calibrated in SaaS systems. The independent calibration experiment and fixed loadings provide information that a deployment would have to obtain separately.

Incident prediction gives a more restrained result. Twin's Brier score is \TwinBrier{}, compared with \EWMABrier{} for EWMA and \NoSurveyBrier{} without surveys. These small differences do not demonstrate a meaningful reduction in operational incidents. Figure~\ref{fig:decision}(a) shows broadly similar calibration curves. A latent coordinate is an intermediate target, and the known simulator outcome law still includes irreducible randomness. Optimizing recovery of that coordinate can therefore deliver a substantial RMSE gain with little predictive or decision advantage. We make no statistically supported claim of incident-forecast superiority.

\subsection{Missingness, misspecification, and forecast intervals}
Figure~\ref{fig:robustness}(a) reports paired stress results. Twin's RMSE is \OutageRMSE{} with extensive loss, \MnarRMSE{} under informative loss, and \ShiftRMSE{} after the noise increase. Its latent coverage changes to \OutageStateCoverage{}\%, \MnarStateCoverage{}\%, and \ShiftStateCoverage{}\%, respectively. Informative loss is not explicitly modeled, and the fitted observation covariance is not recalibrated after the shift. Thus reference coverage is not a reliable forecast of performance when measurement conditions change.

Table~\ref{tab:intervals} separates future noisy telemetry from latent-state intervals. Under increased noise, Gaussian observation coverage is \ShiftGaussianCoverage{}\%; ACI and adapted PID reach \ShiftACICoverage{}\% and \ShiftPIDCoverage{}\%. The corresponding interval scores are \ShiftGaussianScore{}, \ShiftACIScore{}, and \ShiftPIDScore{}. Adaptive methods improve this particular shift response by widening intervals, but neither restores an assured finite-window 90\% level. Figure~\ref{fig:robustness}(b) also shows the delay after the change. The rolling, ACI, and PID results are close; the experiment does not establish a controller winner. All outcomes are conditioned on targets actually observed, so strong informative loss can further separate this measured coverage from coverage of missing events.

\begin{table}[t]
\caption{One-step observed-telemetry intervals, averaged over seeds. Coverage is compared with 90\%; width and proper interval score should be considered jointly. ACI and PID are recent adapted comparators for this task.}\label{tab:intervals}
\centering\small\begin{tabular}{llrrr}
\toprule
Scenario & Interval & Cov. (\%) & Width & Score $\downarrow$ \\
\midrule
reference & Gaussian & 89.7 & 1.090 & 1.381 \\
 & Rolling & 91.1 & 1.169 & 1.426 \\
 & ACI & 91.1 & 1.173 & 1.430 \\
 & PID (adapted) & 91.0 & 1.167 & 1.427 \\
outage & Gaussian & 89.3 & 1.112 & 1.414 \\
 & Rolling & 91.7 & 1.249 & 1.479 \\
 & ACI & 91.7 & 1.251 & 1.480 \\
 & PID (adapted) & 91.7 & 1.248 & 1.479 \\
mnar & Gaussian & 89.6 & 1.099 & 1.387 \\
 & Rolling & 91.1 & 1.194 & 1.439 \\
 & ACI & 91.2 & 1.200 & 1.442 \\
 & PID (adapted) & 91.0 & 1.193 & 1.439 \\
noise shift & Gaussian & 75.4 & 1.090 & 2.452 \\
 & Rolling & 83.5 & 1.354 & 2.188 \\
 & ACI & 84.1 & 1.389 & 2.185 \\
 & PID (adapted) & 83.5 & 1.353 & 2.188 \\
\bottomrule
\end{tabular}

\end{table}

The reference Gaussian score can be better than wider rolling intervals even when the latter have higher coverage. This illustrates why coverage alone is insufficient. Controller adaptation also cannot correct an invalid capability mapping: its target is the next available sensor value, not the organizational meaning assigned to that value. Neither these interval comparisons nor their coverage rates should be read as a state-of-the-art maturity-assessment benchmark.

\subsection{Intervention screening and uncertainty cost}
Figure~\ref{fig:pilot}(a) and Table~\ref{tab:effects} summarize randomized synthetic contrasts. The intervals in the table are the proportion of within-seed bootstrap intervals that include the configured shock; the figure instead shows uncertainty in the across-seed mean estimate. These are different quantities. With six training organizations per arm, effects remain uncertain even under known randomization and persistence. In field data, concurrent changes and endogenous rollout would require a different causal design.

\begin{table}[t]
\caption{Randomized synthetic shock estimates. Sample standard deviation is across seeds. Coverage is based on ten within-seed 95\% organization-bootstrap intervals; it is a coarse diagnostic, not a formal calibration guarantee.}\label{tab:effects}
\centering\small\begin{tabular}{lrrr}
\toprule
Action & True shock & Estimated shock & 95\% CI covers \\
\midrule
Delivery & 0.35 & 0.317 $\pm$ 0.136 & 90\% \\
Recovery & 0.45 & 0.409 $\pm$ 0.098 & 100\% \\
Security & 0.25 & 0.173 $\pm$ 0.153 & 80\% \\
\bottomrule
\end{tabular}

\end{table}

Table~\ref{tab:policies} and Fig.~\ref{fig:decision}(b) show that the mean-utility policy has regret \MeanRegret{}, compared with \RandomRegret{} for random utility ordering. The conservative policy has regret \ConservativeRegret{} and defers \ConservativeDeferred{} of twelve organizations on average. Its caution sacrifices benefit in a simulator where all configured shocks are helpful. This is an informative limitation of uncertainty-based screening rather than a reason to suppress the result. The oracle is an unattainable simulator benchmark, not a reproduced competing method. Its value also depends on the synthetic outcome law and budget.

\begin{table}[t]
\caption{Instantaneous simulator policy outcomes. Benefit is the summed expected incident-risk reduction over twelve test organizations, not an observed count of prevented incidents. Cost is in synthetic units and cannot exceed 48.}\label{tab:policies}
\centering\small\begin{tabular}{lrrrr}
\toprule
Policy & Benefit $\uparrow$ & Regret $\downarrow$ & Cost & Deferred \\
\midrule
MeanUtility & 0.565 & 0.115 & 47.4 & 3.8 \\
Conservative & 0.387 & 0.293 & 33.3 & 6.3 \\
Random & 0.432 & 0.248 & 48.0 & 1.8 \\
Oracle & 0.680 & 0.000 & 48.0 & 6.0 \\
\bottomrule
\end{tabular}

\end{table}

\subsection{Sensitivity, execution cost, and real-data boundary}
Figure~\ref{fig:sensitivity}(a) shows preintervention predictive likelihood across process variances, rather than selecting a parameter by test RMSE. The small three-value search is auditable but cannot establish broad optimality. Figure~\ref{fig:sensitivity}(b) and Table~\ref{tab:scaling} show approximately linear filtering cost across independent organizations at fixed state dimension. The complete simulation execution takes \ElapsedSeconds{} seconds and reaches \PeakRSS{} MiB process peak RSS in the recorded environment. Scaling timings include allocation tracing and are not directly comparable to untraced main-run timings. One run at each size cannot support a timing significance claim or a service-level latency guarantee.

\begin{table}[t]
\caption{Single-run scaling with Python allocation tracing enabled. Peak allocations exclude native-library memory; the complete process RSS is recorded separately.}\label{tab:scaling}
\centering\small\begin{tabular}{rrr}
\toprule
Organizations & Seconds & Python peak (MiB) \\
\midrule
3 & 0.073 & 0.059 \\
6 & 0.145 & 0.118 \\
12 & 0.282 & 0.235 \\
24 & 0.566 & 0.469 \\
48 & 1.144 & 0.937 \\
\bottomrule
\end{tabular}

\end{table}

The historical CI pilot includes \CITestRows{} test builds from \CITestProjects{} projects. Table~\ref{tab:ci-results} and Fig.~\ref{fig:pilot}(b) show that prior build outcomes support useful prediction under both project and time shifts. Logistic regression's Brier score is \CILogisticBrier{}, while the random-forest mean is \CIRandomForestBrier{}. Their numerical difference is small relative to the project-bootstrap intervals. The pilot supports a reproducible data/availability workflow; it does not favor a complex latent model and does not test the full Twin estimator, because the required survey and calibrated multiview capability observations are absent.

\begin{table}[t]
\caption{Historical CI feasibility results. Point estimates average five algorithm seeds on one fixed sample; project-bootstrap bands use first-seed predictions and are not confidence intervals for that seed average. Seed-specific results are retained. No SaaS maturity baseline is claimed.}\label{tab:ci-results}
\centering\small\setlength{\tabcolsep}{4pt}\begin{tabular}{lrrr}
\toprule
Predictor & Brier $\downarrow$ & Cluster 95\% interval & AUROC $\uparrow$ \\
\midrule
Prevalence & 0.1704 & [0.1545, 0.1898] & 0.5000 \\
Trailing rate & 0.1341 & [0.1177, 0.1512] & 0.7275 \\
Logistic & 0.1289 & [0.1131, 0.1452] & 0.7392 \\
Random forest & 0.1293 & [0.1144, 0.1449] & 0.7331 \\
Histogram boosting & 0.1295 & [0.1144, 0.1444] & 0.7332 \\
\bottomrule
\end{tabular}

\end{table}

Failure prevalence decreases between training and testing, and public projects are not a sample of SaaS organizations. Retrospective activity thresholds also condition which projects enter the pilot. These factors limit interpretation even when direct temporal leakage is controlled. The experiment therefore neither validates synthetic capability recovery on real data nor supports transferring the measured probability scores to production incidents.

\section{Threats to Validity and Readiness}\label{sec:limitations}
\subsection{Construct and internal validity}
The principal limitation is that capability truth is simulated. The three labels, unit loadings, survey offset law, intervention shocks, and incident mechanism were chosen by the study designer. Good recovery demonstrates behavior within those assumptions, not a valid organizational maturity instrument. The estimator also receives a separate calibrated noise sample, a substantive advantage that must be matched by instrument studies before field use. Conditional intervals omit parameter, mapping, and causal-model uncertainty. No expert agreement, test--retest reliability of a real survey, or measurement invariance across SaaS contexts was measured.

The generator differs from the fitted state transition, but retains the same broad linear-Gaussian observation family. Noise shifts and informative loss stress that family only partially. Heavy tails, changing loadings, coordinated source manipulation, delayed reporting, and intervention-induced telemetry adoption remain untested. The no-survey, diagonal-covariance, and no-dynamics comparisons isolate specific mechanisms with a shared process variance; they are not independently retuned alternatives. Validation likelihood also evaluates only one source per capability. The protocol is therefore a bounded sensitivity study, not an exhaustive robustness claim.

Unit checks enforce future-observation invariance, positive covariance under complete missingness, dependence-aware precision, shared latent scenarios, exact small-budget optimization, and correct conformal order statistics. A separate computational review found and corrected an interval-rank error and a policy-horizon mismatch before the retained runs. This strengthens implementation confidence while illustrating why generated mathematics and code require independent review.

\subsection{External and statistical validity}
Ten seeds are ten simulated cohorts, not ten industrial replications. Weekly observations within an organization are dependent and are not counted as independent samples for the main contrasts. Sign-flip inference relies on its stated symmetry or exchangeability null. Bootstrap intervals with six organizations per treatment arm and ten seed-level contrasts are approximate, and their tail precision is limited. Decision utility combines state draws with frequentist effect resampling; calling it a fully Bayesian causal posterior would overstate what was computed.

The historical pilot uses records from 2011--2015 in a later snapshot, a different CI ecosystem from current SaaS delivery. Completion time is inferred from reported duration, and forks or shared code lineages are not screened. Public-project identifiers are hashed in the processed release to minimize unnecessary identifiers, but hashing is pseudonymization rather than a confidentiality guarantee. Dataset attribution and source checksums are retained. No individual engineering performance or employee ranking should be inferred from these project outcomes.

The model does not assess comprehensive SaaS business maturity, tenancy isolation, customer value, security effectiveness, or monetary savings. The synthetic security coordinate has no real scan or vulnerability labels. Real deployment would require prospective mapping validation, source calibration, availability semantics, privacy governance, and human review of recommendations. At least three partner organizations would provide a starting point for a longitudinal feasibility study, but three clusters alone do not ensure adequately powered causal inference.

\subsection{Submission and novelty boundary}
The literature audit includes established continuous assessment and Bayesian delivery systems. We claim an inspectable prototype and evaluation protocol, without asserting first use, unique integration, or domain state-of-the-art performance. ACI and PID are recent algorithmic observation-interval comparators, not faithful reproductions of the closest SaaS/DevOps systems. The requested comparison with two current domain state-of-the-art methods therefore remains unmet.

IEEE Transactions on Services Computing is the conditional target on scope grounds. The audit records SCImago 2025 Q1 evidence in Computer Science Applications and other specified categories; this is neither a 2026 ranking nor a JCR classification. The exact journal submission cap, fees, anonymization options, and final author responsibilities still require verification at submission. The package is a reproducible research draft. Its submission checklist remains blocked by field construct validation, a convincing methodological novelty case, and matched domain comparisons. Formatting, reference quotas, and a compiled PDF cannot substitute for those scientific requirements.

\section{Conclusion}\label{sec:conclusion}
This study examined how a continuous technical capability estimate should respond to biased surveys, dependent telemetry, and missing observations. SAAS-MATURITY-TWIN combines an anchored Gaussian state model, a persistent survey offset, covariance aware updates, and conditional uncertainty. A separate action screening procedure uses randomized synthetic contrasts and an explicit budget, while preserving the distinction between inferred states and intervention evidence. Across ten simulated cohorts, reference latent error was \TwinRMSE{}, compared with \EWMARMSE{} for tuned moving averages. Ninety percent intervals covered \TwinCoverage{} percent of states. These improvements did not translate into substantial incident forecasting gains. Increased noise reduced calibration, and recent adaptive interval comparators improved coverage without establishing reliable local guarantees. Conservative intervention screening deferred useful actions and incurred greater simulator regret than mean utility selection. Historical TravisTorrent forecasting showed that simple logistic regression remained competitive with larger models under project and temporal separation. The main limitation is the absence of validated longitudinal SaaS capability measurements and real intervention outcomes. Independent calibration, fixed mappings, and synthetic structure constrain practical interpretation. Future work should prospectively validate measurement invariance and outcome relationships across partner organizations, then evaluate appropriately identified interventions. The present package supports reproducible investigation, while remaining a research draft requiring field evidence.

\bibliographystyle{IEEEtran}
\bibliography{references}
\end{document}